\documentclass{article}
\usepackage{spconf,amsmath,graphicx,url,mathrsfs,amssymb,hyperref,booktabs,multirow}

\title{Quantum-Inspired Vision: 
	Leveraging Wave-Particle \\Duality for Low-Illumination Enhancement}
\name{Yiquan Gao\thanks{See \protect\url{https://github.com/StudioYG/DRU} for further details.}}
\address{Heriot-Watt University}

\begin{document}
%
\maketitle
\begin{abstract}
	This study provides a theoretical expansion of the recent Data Relativistic Uncertainty (DRU) framework by formalizing a physics-to-AI paradigm for image enhancement. By modeling images as probabilistic wave functions rather than deterministic states, the paradigm explicitly integrates wave-particle duality to illustrate the system flow of how DRU leverages the intrinsic physical uncertainty of light, a dimension requiring further theoretical discussion. Consequently, this paradigm provides a rigorous  Explainable AI (XAI) approach that enhances the interpretability of how DRU mitigates illumination bias and maintains robustness against data noise.
\end{abstract}
\begin{keywords}
Data relativistic uncertainty (DRU), image enhancement, physics-to-AI paradigm, wave-particle duality
\end{keywords}
\section{Introduction}
\label{sec:intro}

The rapid advancement of AI in image enhancement has often come at the cost of ignoring the fundamental physical properties of light. Most existing models rely on deterministic assumptions during training, treating illumination uncertainty as fixed values. However, in complex environments, this deterministic view fails to account for illumination variability, frequently resulting in over- or under-enhancement.

To address this, this work extends the Data Relativistic Uncertainty (DRU) framework \cite{gao2025data} through a \textit{theoretical expansion and interpretive paradigm study} for image enhancement. While the extensive empirical benchmarking, dataset validations, and baseline state-of-the-art (SOTA) architectural comparisons have been thoroughly established in \cite{gao2025data}, this study specifically targets a critical Explainable AI (XAI) gap in DRU. A physics-to-AI paradigm is introduced to treat image samples as probabilistic wave functions, moving beyond the framework’s original empirical scope to provide a formal theoretical discussion connecting quantum concepts and DRU. By leveraging wave-particle duality~\cite{selleri1992wave}, it defines a system flow where each sample exists in a superposition of bright and dark states, explaining the underlying mechanism of \textit{why} and \textit{how} the DRU framework operates robustly. Within this paradigm, the loss function calculation exhibits a structural mapping analogous to the dynamic evolution of a definite state observation or particle-like measurement, allowing the model to interpretably calibrate each sample’s contribution based on state probabilities.

By modeling image samples as probabilistic wave functions rather than deterministic states, this paradigm provides the exact theoretical mechanism for how DRU formalizes inherent illumination uncertainty. It establishes a physics-informed foundation that enhances the expressivity of enhancement networks in structural fidelity and aesthetic quality, effectively bridging the gap between AI Explainability and the underlying Wave-Particle Duality. 

The primary contributions of this work are summarized as follows:
\begin{itemize}
	\item A rigorous physics-to-AI paradigm is established by modeling image restoration via a \textit{formal structural mapping} to a probabilistic state collapse, providing the theoretical foundation for the Data Relativistic Uncertainty (DRU) framework.
	\item A unified perception-to-aesthetic metric is proposed to reconcile divergent optimization directions of BRISQUE and NIMA, serving as the analytical tool for the empirical quantification of coupled error components.
	\item The proposed paradigm underpins both the quantitative comparisons and transition-based error analysis, elucidating the theoretical roots of the DRU mechanism's robustness.
	
\end{itemize}

\section{Related Works}
\textbf{Physics-Driven Learning.} Physics-driven learning incorporates physical priors into data-driven models to enhance generalization and data efficiency. Physics-Informed Neural Networks (PINNs)~\cite{cai2021physics} embed governing equations into the loss, while neural differential equation models~\cite{chen2018neural,bilovs2021neural} (e.g., Neural ODEs) provide continuous-time dynamics. Symmetry-preserving architectures~\cite{favoni2025symmetry} enforce physical structure, and operator learning methods, such as Fourier Neural Operators~\cite{li2020fourier}, learn mappings between function spaces. Hybrid models~\cite{zou2024hybrid2} further combine mechanistic and neural components to capture system dynamics. These approaches primarily focus on structural or macroscopic constraints from physics. However, this study establishes a rigorous theoretical basis for the Data Relativistic Uncertainty (DRU) framework~\cite{gao2025data} grounded in the principles of Wave-Particle Duality.

\noindent\textbf{Wave Particle Duality and State Collapse.} Wave Particle Duality~\cite{selleri1992wave} characterizes quantum systems as superpositions over multiple states, represented by a wave function~\cite{bassi2013wave}. Measurement induces state collapse, projecting the superposed state onto a single eigenstate~\cite{orefice2017dynamics}. This interplay between superposition and collapse provides a compact perspective for modeling uncertainty and selective state resolution in learning systems.

\begin{figure*}[t]
	\centering
	\includegraphics[width=\linewidth]{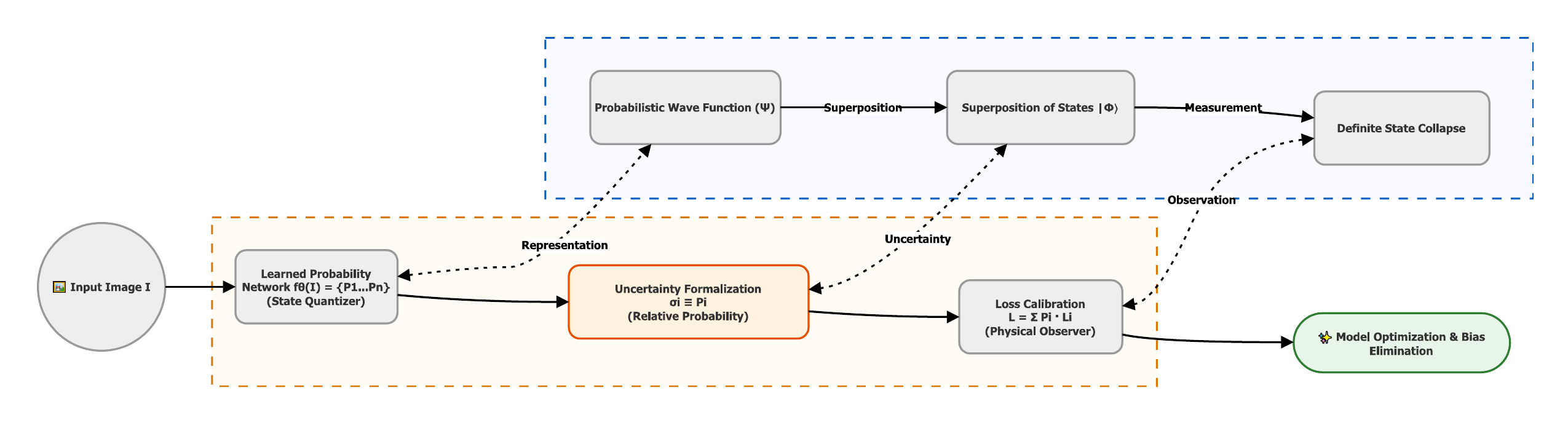}
	\caption{Theoretical flow of the Physics-to-AI paradigm. The process flows from wave function quantization ($f_\theta(I) \rightarrow \Psi$) to linear superposition ($|\Phi\rangle = \sum \sqrt{P_i} |s_i\rangle$). It concludes with a definite state observation via loss calibration ($\mathscr{L} = \sum P_i \cdot \ell_i$), effectively resolving illumination bias during model optimization.}
	\label{fig:paradigm}
\end{figure*}

\section{Theoretical Formalization of the Physics-to-AI Paradigm}

By characterizing the behavior of DRU as a general system flow in Fig.~\ref{fig:paradigm}, a physics-to-AI paradigm bridging Wave-Particle Duality and DRU is proposed as follows:

\subsection{Probabilistic Wave Function Representation ($\Psi$)}
 To operationalize the wave nature of light, the paradigm introduces an architecture-agnostic Learned Probability Network $f_\theta(\cdot)$ (e.g., CNNs or Transformers) acting as a specialized state quantizer. This network maps the input image $I$ into a Wave Function ($\Psi$) space where bright and dark illumination states coexist:
\begin{equation}
	\Psi(I) = f_\theta(I) = \{P_1, P_2, \dots, P_n\}
\end{equation}
This probabilistic representation captures inherent physical uncertainties ignored by deterministic models. Here, $n$ is the number of illumination states, depending on the configuration of enhancement networks. The network $f_\theta(\cdot)$ outputs Relativistic Probability (RP) values $P$, formalizing the physical likelihood of each potential illumination state.

\subsection{Superposition and Illumination Uncertainty}
 Within this $\Psi$ representation, each sample exists in a superposition of states $|\Phi\rangle$, allowing a single input to be simultaneously associated with multiple potential illumination levels $|s_i\rangle$ during training. This superposition is formalized as a linear combination:
\begin{equation}
	|\Phi\rangle = \sum_{i=1}^{n} \sqrt{P_i}\, |s_i\rangle
	\quad \text{subject to} \quad
	P_i \ge 0,\ \sum_{i=1}^{n} P_i = 1
\end{equation}
In this formulation, the global state $|\Phi\rangle$ is defined as a linear superposition where the probability of observing each base state $|s_i\rangle$ is given by the relativistic probability $P_i$. This ensures that the contribution of each illumination state to the training objective is directly proportional to its inherent physical uncertainty, as quantified by the RP values. 
Crucially, the RP value $P_i$ assigned to each state serves as a direct quantification of the physical uncertainty $\sigma_i$ inherent to that state (i.e., $\sigma_i \equiv P_i$). Rather than being an external parameter, uncertainty is treated as an intrinsic property embedded within the state distribution. This representation allows the optimal learning direction to remain fluid until resolved into a definitive signal during the measurement phase.

\subsection{Loss Analogous to Definite State Observation}
Mirroring the particle nature of light, the loss function calculation is formalized via a structural mapping to a state observation process. In this paradigm, the loss function acts through an interpretive analogy to a physical observer; during optimization, the learning process is modeled as the mapping of the probabilistic Wave Function $\Psi(I)$, collapsing from its superposition of states $|\Phi\rangle$ into a localized particle-like state $|s_i\rangle$ with an empirical probability $P_i$. To operationalize this mechanism, the DRU framework performs $\mathscr{L}$ Calibration by formulating the total objective $\mathscr{L}$ as:
\begin{equation}
	\mathscr{L} = \sum_{i=1}^{n} P_i \cdot \ell(I, s_i)
\end{equation}
where $I$ is the input image, $s_i$ denotes the $i$-th potential illumination state (e.g., bright or dark), and $\ell(I, s_i)$ represents the task-specific loss for that state. The term $P_i$ signifies the RP values derived from $\Psi(I)$, which serve as the direct quantification of the intrinsic physical uncertainty $\sigma_i$ for each state $s_i$ (where $\sigma_i \equiv P_i$). By weighting the gradient with these values, the framework interpretably calibrates each sample’s contribution to the learning process. This duality-driven mechanism explicitly models illumination uncertainty to eliminate Illumination Bias, effectively preventing the over- or under-enhancement issues prevalent in deterministic models. Crucially, unlike traditional methods optimizing conditional variance via maximum likelihood estimation, modeling image samples as superposition states allows the low probability amplitudes of noisy samples to mathematically suppress the influence of data noise during this state evaluation phase, ensuring fluid and robust optimization paths.

\section{Experiments}

\begin{table*}[t]
	\centering
	\setlength{\tabcolsep}{12pt}
	\caption{Quantitative comparison (BRISQUE $\downarrow$~\cite{mittal2012no} and NIMA $\uparrow$~\cite{talebi2018nima}) between DRU-based and deterministic models. The results highlight the behavioral divergence stemming from the DRU mechanism across different data scenarios. The upper block contains deterministic models, while the lower block showcases DRU variants, where Arch-Backbone denotes the enhancement architecture paired with its probability network.
	}
	\label{tab:comparison}
	\small
	\begin{tabular}{l c c c c} 
		\toprule
		\multirow{2}{*}{\textbf{Model Type}} & \multicolumn{2}{c}{\textbf{Standard ASD}} & \multicolumn{2}{c}{\textbf{Noisy ASD (Label Noise)}} \\
		\cmidrule(r){2-3} \cmidrule(l){4-5}
		& BRISQUE $\downarrow$ & NIMA $\uparrow$ & BRISQUE $\downarrow$ & NIMA $\uparrow$ \\
		\midrule
		SCI~\cite{ma2022toward} &  27.87 &  4.68 & -- & -- \\
		ZeroDCE++~\cite{li2021learning} &   28.40 &  4.49 & -- & -- \\
		RUAS~\cite{liu2021retinex} & 49.13 &  4.18 & -- & -- \\
		EnGAN~\cite{jiang2021engan} &  27.28 & 4.75 &  29.15 & 4.80 \\
		\midrule
		RUAS-ResNet18 & 34.62 &  4.28 & -- & -- \\
		\textbf{EnGAN-ResNet18} & \textbf{25.97} & \textbf{ 4.76} & \textbf{26.83} & \textbf{4.80} \\
		\bottomrule
	\end{tabular}
\end{table*}

To analyze and interpret the behavioral divergence between the DRU framework and deterministic models, two variants of the unpaired Anime Scenery Dataset (ASD) from \cite{gao2025data} are employed: a standard version and a noisy version containing misclassified samples in the training set, both using an identical test set. Four modern enhancement networks are adopted, with implementation details following their original experimental protocols~\cite{ma2022toward,li2021learning,liu2021retinex,jiang2021engan}. This dual-dataset approach serves as a controlled environment to demonstrate how the proposed paradigm enables the DRU to maintain robustness against data noise, while deterministic models suffer from illumination bias exacerbated by label noise  due to their rigid state assumptions. 

Table \ref{tab:comparison} shows that DRU variants consistently outperform deterministic counterparts. Under standard ASD, EnGAN-ResNet18 achieves state-of-the-art performance (BRISQUE: 25.97, NIMA: 4.76), while RUAS-ResNet18 demonstrates substantial metric improvements. Notably, EnGAN-ResNet18 remains robust under noise, even surpassing deterministic models in standard settings. This confirms the paradigm's practicality in stabilizing and boosting enhancement networks amidst uncertainty.

The evolution from EnGAN to EnGAN-ResNet18 reveals a behavioral shift fully accounted for by the theoretical paradigm. While the performance of EnGAN is constrained by a deterministic assumption (fixing $P_i=1.0$) and further degraded by label noise, the DRU version exhibits unique resilience by maintaining a superior balance between perceived distortion and aesthetic quality from standard to noisy conditions. This reflects a transition from simplistic label-fitting toward the pursuit of physical consistency, effectively overcoming the illumination biases inherent in rigid, deterministic models, while preventing their exacerbation by label noise.

\begin{table*}[t]
	\centering
	\caption{Decoupling Analysis of Error Components. The magnitudes represent the performance gap induced by specific influencing factors. `Noisy' indicates Noisy ASD; otherwise Standard ASD is default.}
	\label{tab:error_decomposition}
	\begin{tabular}{@{}llc@{}}
		\toprule
		\textbf{Error Component} & \textbf{Analysis Transition ($1 \rightarrow 2$)} & \textbf{Calculated Error} $\epsilon$ $\downarrow$  \\ \midrule
		Coupled Error $\epsilon_{\text{couple}}$    & EnGAN-ResNet18 $\rightarrow$ Noisy EnGAN & 0.62 \\
		Illumination Bias $\epsilon_i$        & EnGAN-ResNet18 $\rightarrow$ EnGAN & 0.29 \\
		Label Noise $\epsilon_l$              & EnGAN $\rightarrow$ Noisy EnGAN & 0.33 \\ 
		
		Mitigated Label Noise $\epsilon_{mit}$    & EnGAN-ResNet18 $\rightarrow$ Noisy EnGAN-ResNet18 & 0.13 \\
		\bottomrule
	\end{tabular}
\end{table*}

The essence of this observed robustness is rooted in Wave-Particle Duality. Within this paradigm, the DRU treats the input not as a singular deterministic value, but as a superposition of states $|\Phi\rangle$ encompassing multiple illumination states. Here, the loss function acts as a physical observer: during optimization, the wave function $\Psi(I)$ collapses from its superposition into a localized particle state $|s_i\rangle$ with probability $P_i$. Since noisy samples occupy minimal probability amplitudes, their capacity to exacerbate the illumination bias inherent in deterministic models is naturally suppressed during the collapse. By weighting the loss according to these amplitudes, the DRU shifts the optimization focus toward physically consistent states, preventing noise from amplifying the biases enforced by rigid deterministic assumptions. Consequently, the model favors a collapse into the most physically plausible state, providing a rigorous interpretation for harnessing categorical uncertainty through probabilistic reasoning.

The paradigm allows for the decomposition of the coupled error into intrinsic and external factors, denoted as $\epsilon_{\text{couple}} = \epsilon_i \otimes \epsilon_l$, where $\otimes$ represents the interaction operator between illumination bias $\epsilon_i$ (intrinsic) and label noise $\epsilon_l$ (external). To empirically quantify these components, the error $\epsilon$ is defined based on the analysis transition ($1 \rightarrow 2$) as follows:

\begin{equation}
	\epsilon = \frac{\text{BRISQUE}_2}{\text{NIMA}_2} - \frac{\text{BRISQUE}_1}{\text{NIMA}_1}
\end{equation}
where the subscripts $1$ and $2$ represent the \textit{initial} and \textit{terminal} models of each transition specified in Table \ref{tab:error_decomposition}. This proposed equation harmonizes the opposing optimization directions of BRISQUE and NIMA into a unified descending  perception-to-aesthetic metric, effectively capturing the performance shift attributed to each constituent error factor. Subsequent empirical validation confirms that this formulation is numerically justified, stable, and entirely free of scaling anomalies during analytical decoupling.

Table \ref{tab:error_decomposition} illustrates the paradigm's applicability in decomposing coupled error sources regarding EnGAN. By guiding the analysis of model transitions, the paradigm isolates the error induced by rigid illumination bias ($P_i=1.0$) at $\epsilon_i=0.29$ and the additional degradation from label noise at $\epsilon_l=0.33$. These results indicate that while the deterministic assumption creates a baseline error, stochastic interference from noisy data plays a more dominant role in performance degradation. Notably, the label noise error is reduced by 0.2 via the quantum mechanism of DRU. This demonstrates that the theoretical paradigm provides the necessary clarity to disentangle intertwined factors into identifiable individual impacts.

\section{Conclusion}

In conclusion, this work establishes a rigorous physics-to-AI paradigm that expands the theoretical foundation of the Data Relativistic Uncertainty (DRU) framework. By formulating image enhancement through the lens of wave-particle duality, it moves beyond deterministic state assumptions to model images as probabilistic wave functions. This transition effectively redefines the restoration process via a formal structural mapping to a state collapse process, where the intrinsic uncertainty of light is leveraged to suppress illumination biases. Experimental analysis shows that the DRU framework avoids simplistic label-fitting by prioritizing physical consistency, a shift that explains its proven robustness against label noise. Ultimately, this paradigm provides an interpretive and robust pathway for stabilizing enhancement networks amidst the inherent uncertainty of low-illumination data.

\section*{Acknowledgments}
The author would like to acknowledge the enduring principles of quantum mechanics, particularly the concept of wave-particle duality, which provided the fundamental inspiration for the theoretical paradigm explored in this work.

\bibliographystyle{IEEEbib}
\bibliography{strings,refs}

\end{document}